\input harvmac

\def\lf{16\pi^2}

\def\ha{{1\over2}}
\def\half{{\textstyle{1\over2}}} 
\def\ga{\gamma}
\def\de{\delta}

\def \la{\lambda}

\def\sy{supersymmetry}
\def\sic{supersymmetric}
   
\def\ssm{supersymmetric standard model}
\def\sm{standard model}

\def\pa{\partial}
\def\321{$SU_3\otimes\ SU_2\otimes\ U_1$}

\def\npb{{Nucl.\ Phys.\ }{\bf B}}

\def\plb{{Phys.\ Lett.\ }{\bf B}}

\def\prd{{Phys.\ Rev.\ }{\bf D}}

\def\msbar{${\overline{\hbox{MS}}}$}

\def \in{\leftskip = 40 pt\rightskip = 40pt}
\def \out{\leftskip = 0 pt\rightskip = 0pt}

\def\llf{(16\pi^2)^2}
\def\lllf{(16\pi^2)^3}
\def\llllf{(16\pi^2)^4}
{\nopagenumbers
\line{\hfil LTH 368}
\line{\hfil hep-ph/9603386}
\vskip .5in
\centerline{\titlefont $N=1$ supersymmetry and the 
three loop}
\centerline{\titlefont anomalous 
dimension  for the chiral superfield}
\vskip 1in
\centerline{\bf I.~Jack, D.R.T.~Jones and C.G.~North}
\bigskip
\centerline{\it DAMTP, University of Liverpool, Liverpool L69 3BX, U.K.}
\vskip .3in

We calculate the three loop anomalous dimension for a general 
$N=1$ supersymmetric gauge theory. The result is used to probe the 
possible existence of renormalisation invariant relationships 
between the Yukawa and gauge couplings.

\Date{March 1996}}

\newsec{Introduction}

In a recent paper \ref\jjb{I.~Jack and D.R.T.~Jones, \plb 349 (1995)
294} we showed that in certain \sic\ theories, it is possible to impose
a  relation  among the dimensionless  couplings  (which we shall call
generically the $P={1\over3}Q$ condition), which is preserved under
renormalisation through at least two loops. In these theories, then if  
$P={1\over3}Q$ is not actually imposed, it nevertheless  represents a
renormalisation-group (RG) fixed point of the coupling evolution: 
in many theories an infra--red stable one.     In addition we 
showed that  in such theories the  soft \sy-breaking couplings may 
take  (or approach in the infra--red) a particular universal form which 
is also preserved by RG evolution.

We have pursued the phenomenological consequences of these ideas
elsewhere\ref\jjra{I.~Jack, D.R.T.~Jones and K.L.~Roberts, \npb 455
(1995) 83}. These are valid whether $P={1\over3}Q$ is an infra--red 
phenomenon
\ref\fjj{P.~Ferreira, I.~Jack and D.R.T.~Jones,\plb 357 (1995) 359}
\ref\lanz{M.~Lanzagorta and G.G.~Ross, \plb 364 (1995) 163}
or a consequence of some fundamental symmetry. 
In this paper we ask whether the
special properties of theories with $P={1\over3}Q$ persist to higher orders.
We postulate an all-orders relation between the gauge $\beta$-function and the
anomalous dimension of the chiral supermultiplet for a $P={1\over3}Q$ theory,
and then investigate its validity. 
For this purpose it is sufficient to
calculate the three-loop  contribution $\ga^{(3)}$  to the anomalous
dimension of the chiral supermultiplet  in a general  $N=1$
supersymmetric gauge theory.  This calculation is  an extension to the
existing one of Parkes\ref\parkes{A.J.~Parkes, \plb 156 (1985) 73}, 
where he calculated $\ga^{(3)}$ for a non--abelian one--loop finite
theory,  in turn generalising the existing result of 
Refs.~\ref\grs{M.~Grisaru, M.~Rocek and W.~Siegel, \npb 183 (1981) 141} and 
\ref\cz{W.E.~Caswell and D.~Zanon, \npb 182 (1981) 125} for the $N=4$
theory.

Our result will have other uses; we may, for example, calculate the 
Yukawa $\beta$-functions for the \sic\ standard model and investigate the 
domain of perturbative believability for the $t$-quark Yukawa coupling.  

\newsec{The ${\bf P={1\over3}Q}$ condition}

The Lagrangian $L_{\rm SUSY} (W)$ for a $N=1$ \sic\ theory 
is defined by the superpotential
\eqn\Ea{
W={1\over6}Y^{ijk}\Phi_i\Phi_j\Phi_k+{1\over2}\mu^{ij}\Phi_i\Phi_j. }
$L_{\rm SUSY}$ is the Lagrangian for  the $N=1$ supersymmetric
gauge theory, containing the gauge multiplet ($\lambda$ being the
gaugino) and a chiral superfield $\Phi_i$ with component fields
$\{\phi_i,\psi_i\}$ transforming as a 
representation $R$ of the gauge group $\cal G$. 
We assume that there are no gauge-singlet
fields and that $\cal G$ is simple.

The superpotential $W$ undergoes no infinite renormalisation;  
so that we have, using  minimal subtraction (MS) or 
modified minimal subtraction (\msbar)
\eqn\Ec{
\beta_Y^{ijk}= Y^{p(ij}\ga^{k)}{}_p = 
Y^{ijp}\ga^k{}_p+(k\leftrightarrow i)+(k\leftrightarrow j),}
where $\ga$ is the anomalous dimension for $\Phi$. Note that Eq.~\Ec\ 
depends on use of MS or \msbar\ as well as on a \sic\ regularisation 
method; we shall 
see  later that this relationship between $\beta_Y$ and $\ga$ is
not  preserved by certain coupling constant redefinitions. 

The one-loop results for the gauge coupling $\beta$-function $\beta_g$ and 
for $\ga$ are given by
\eqn\Ed{
\lf\beta_g^{(1)}=g^3Q,\quad\hbox{and}\quad 
\lf\ga^{(1)i}{}_j=P^i{}_j,}
where
\eqna\Ee$$\eqalignno{ 
Q&=T(R)-3C(G),\quad\hbox{and}\quad &\Ee a\cr
P^i{}_j&={1\over2}Y^{ikl}Y_{jkl}-2g^2C(R)^i{}_j. &\Ee b\cr}$$
Here
\eqn\Ef{
T(R)\delta_{AB} = \Tr(R_A R_B),\quad C(G)\delta_{AB} = f_{ACD}f_{BCD} 
\quad\hbox{and}\quad C(R)^i{}_j = (R_A R_A)^i{}_j.}

The two-loop
$\beta$-functions for the dimensionless couplings were calculated in
\break Refs.~\ref\tja{D.R.T.~Jones, \npb87 (1975) 127}%
\nref\pwa{A.J.~Parkes and P.C.~West, \plb138 (1984) 99}%
\nref\pwb{A.J.~Parkes and P.C.~West, \npb256 (1985) 340}%
\nref\west{P.~West, \plb137 (1984) 371}%
--\ref\tjlm{D.R.T. Jones and L. Mezincescu, \plb136 (1984) 242; 
{\it ibid} 138 (1984) 293}: 
\eqna\Au$$\eqalignno{ \llf\beta_g^{(2)}&=2g^5C(G)Q-2g^3r^{-1}C(R)^i{}_jP^j{}_i
&\Au a\cr
\llf\ga^{(2)i}{}_j&=[-Y_{jmn}Y^{mpi}-2g^2C(R)^p{}_j\delta^i{}_n]P^n{}_p+
2g^4C(R)^i{}_jQ,&\Au b\cr}
$$
where $Q$ and $P^i{}_j$ are given by Eq.~\Ee{}, and $r=\delta_{AA}$.

The $P={1\over3}Q$ condition is the requirement that
\eqn\El{
P^i{}_j = {1\over3}g^2Q\delta^i{}_j,}
or equivalently
\eqn\Eld{ \ga^{(1)} = {{\beta_g^{(1)}}\over{3g}}\de^i{}_j,}
and thus amounts to a postulated relation between the 
Yukawa and gauge couplings. 
It is easy to show from Eqs.~\Ec--\Ee{}\ and \Au{}\ that Eq.~\El\ 
corresponds to a fixed point in the evolution of $Y^{ijk} /g$, up to two-loop
order;
in other words, a possible solution to the 
equation 
\eqn\Ela{
\mu{d \over{d\mu}} {{Y^{ijk}}\over g} = 0, }
or equivalently 
\eqn\Elb{
\beta_Y^{ijk}= g^{-1}Y^{ijk}\beta_g.}

While Eq.~\El\ solves Eq.~\Elb\ up to two loops,
it is highly restrictive, and in many 
cases there is no choice of the Yukawa couplings that corresponds to a 
solution of Eq.~\El. The $t$-quark Yukawa 
evolution in  the \sm\ (or indeed in  the \ssm ) 
is a familiar example of this 
type \ref\Pendleton{B.~Pendleton and G.G.~Ross, \plb 98 (1981) 291}. The 
attractive feature of theories that do admit $P = {1\over 3}Q$ is 
that in such theories, the 
soft \sy-breaking couplings also 
have fixed points which correspond to  the commonly 
assumed universal form\jjb--\fjj. 

In general, the solution to Eq.~\Elb\ (to all orders) 
would be a power series of the form
\eqn\Elba{Y^{ijk}=ga_1^{ijk}+g^3a_3^{ijk}+g^5a_5^{ijk}+\ldots,}
where $a_1$, $a_3$, $a_5$ etc are constant tensors.
What we have shown is that $a_3^{ijk}=0$ if $Y^{ijk}=ga_1^{ijk}$ satisfies
Eq.~\El, in other words if
\eqn\Elbb{{1\over2}a_1^{ikl}a_{1jkl}-2C(R)^i{}_j={1\over3}Q\delta^i{}_j.}
Our investigation in this paper is motivated by 
the question of whether Eq.~\El\
corresponds to a fixed point of $Y^{ijk} /g$ to all orders, or in other words 
whether in Eq.~\Elba, we have $a_5=a_7=\ldots=0$ 
in addition to $a_3=0$, so that
\eqn\Elc{Y^{ijk} = ga_1^{ijk}}
gives a fixed point of $Y^{ijk} /g$ to all orders.

In a situation of physical interest such as a supersymmetric grand 
unified theory, the symmetry of some underlying theory might guarantee
that Eq.~\El\ was satisfied at energies of order the Planck scale. If Eq.~\El\
corresponded to an exact fixed point, then it would continue to be exactly 
satisfied during the RG evolution. On the other hand, the fixed point might be
approached in the infra-red in the absence of 
any special boundary conditions at the Planck scale. 

Fixed points of $Y^{ijk} /g$ are related to the
coupling constant reduction  (CCR) program pursued in 
Ref.~\ref\zimm{N.-P. Chang, \prd10 (1974) 2706\semi
N.-P. Chang, A. Das and J. Perez-Mercader, \prd 22 (1980) 1829\semi
R. Oehme, K. Sibold and W. Zimmermann, \plb 153 (1985) 142\semi
R. Oehme and W. Zimmermann  Com. Math. Phys 97 (1985) 569\semi
W. Zimmermann, {\it ibid} 97 (1985) 211\semi
R. Oehme,  Prog. Theor. Phys. Suppl. 86 (1986) 215\semi
R. Oehme, hep-th/9511006}. 
A fixed point of the form Eq.~\Elc\ is the simplest possible 
realisation of CCR.
In general, CCR proceeds by the 
assumption that in a many-coupling theory, all the couplings 
 may be expressed in terms of one of them (typically the gauge coupling) 
by relations which in our notation would take the form: 
\eqn\Ema{Y^{ijk} = \la^{ijk} (g)}
whence 
\eqn\Emb{\beta_Y^{ijk} = {{d\la^{ijk}}\over{dg}}\beta_g.}
Clearly Eq.~\Elc\ corresponds to the simplest possible result 
for $\la^{ijk}$; in general one might expect 
\eqn\Emc{\la^{ijk}= ga^{ijk} + g^3 b^{ijk} + g^5 c^{ijk}+\ldots,}
where $a^{ijk}$, $b^{ijk}$, $c^{ijk}$ etc are $\mu$-independent constant 
tensors.   
We shall see later that this general situation is equivalent to a fixed point 
for a redefined $Y^{ijk} /g$.
We shall return to a discussion of CCR in the 
$P={1\over3}Q$ case, for which $a^{ijk}=a_1^{ijk}$, later.

The $P={1\over3}Q$ condition itself is RG
invariant at a fixed point of $Y^{ijk}/g$. Indeed, differentiating 
Eq.~\El\ with respect to $\mu$ we obtain 
\eqn\Ele{{1\over2}\left\{\beta_Y^{ikl}Y_{jkl}+Y^{ikl}\beta_{Yjkl}\right\}
-4g\beta_gC(R)^i{}_j={2\over3}g\beta_gQ\delta^i{}_j,}
which is satisfied when Eq.~\Elb\ holds; in other words Eq.~\Elb\ implies that
the $P={1\over3}Q$ condition is RG invariant.  
It is easy to check that the reason Eq.~\Elb\ is satisfied up to two 
loops by couplings satisfying Eq.~\El\ is that Eq.~\El\ also implies
 \eqn\En{ \ga^{(2)i}{}_j = {{\beta_g^{(2)}}\over{3g}}\de^i{}_j,}
which readily follows from Eqs.~\Au{}\ and \Ee{}. This corresponds to 
having $b^{ijk} = 0$ in Eq.~\Emc. 
It is then natural to speculate 
that this relation might be completely general, so that 
\eqn\Ena{ \ga^i{}_j = {{\beta_g}\over{3g}}\de^i{}_j}
to all orders, provided we impose Eq.~\El. It is this hypothesis which we aim 
to check.

At this point we encounter scheme--dependence problems.  It is obvious
without any calculations that Eq.~\Ena\ will not be true  if we use DRED
and MS or \msbar, since we know \parkes\pwb\ that in a two--loop  finite
theory, $\ga^{(3)}$ is non--zero, while $\beta_g^{(3)}$ is zero.   All
is not lost, however, since in such a case $\ga^{(3)}$ may  be
transformed to zero by a coupling constant redefinition (as we shall 
show later). Such redefinitions are equivalent to a change of
renormalisation scheme.   We might have hoped, therefore, that in this
new scheme Eq.~\Ena\ would  hold in the non--finite case.  We will see,
however, that (essentially because $C(R)$ is not in  general
proportional to the unit matrix) it is not possible to achieve 
Eq.~\Ena\ even with arbitrary coupling constant redefinitions.    Even
had we succeeded, however, the significance of the result for the  fixed
point discussion would have  been unclear, since after coupling constant
redefinitions Eq.~\Ec\ no longer  holds (except in the finite case), so
that Eq.~\Ena\ no longer corresponds  to a solution  of Eq.~\Elb.  We
discuss these issues in more detail in section 7. 

\newsec{${\bf\ga}$ and the Instanton--based ${\bf\beta_g}$}

There exists an exact relation between $\ga$ and $\beta_g$, 
derived by Novikov  et al. and based on the instanton calculus
\ref\nov{V.A.~Novikov et al, 
\plb166 (1986) 329}\ref\nova{
V.A.~Novikov et al, 
\npb 277 (1986) 
456}. In our notation it reads: 
\eqn\russa{\beta_g = 
{{g^3}\over{\lf}}\left[ {{Q- 2r^{-1}\Tr\left[\ga C(R)\right]}
\over{1- 2C(G)g^2{(\lf)}^{-1}}}\right].}
It is not entirely clear from Ref.~\nov, \nova\ how general this result for 
$\beta_g$ is. For example, in Ref.~\nova, one sees from the expression given 
for $\ga^{(1)}$ that in this paper $Y^{ijk} =0$ is assumed. In Ref.~\nov, 
however, it is clearly asserted that the result stands for non-zero Yukawa 
couplings.  Also applications of Eq.~\russa\ of which we are aware have been 
to cases such that $C(R)\sim \de^i{}_j$.  
We shall in any case 
assume that Eq.~\russa\ is true for a general theory
with a superpotential. 

In the case $\ga =0$ (no matter fields) Eq.~\russa\ 
was first obtained in Ref.~\ref\jab{D.R.T.~Jones, \plb 123 (1983) 45}; 
see also Ref.~\pwb\ and Ref.~\ref\mtgw{M.T.~Grisaru and P.~West, \npb   
254 (1985) 249}.
Note that 
in Eq.~\russa, the $(n+1)$th loop contribution to $\beta_g$ , 
$\beta_g^{(n+1)}$, is essentially 
determined  by
$\ga^{(n)}$, and that for an $n$-loop finite theory we have 
automatically that $\beta_g^{(n+1)} =0$. Thus Eq.~\russa\ 
is consistent with Refs.~\pwb\ and 
\ref\gmz{M.T.~Grisaru, B.~Milewski and D.~Zanon, \plb 155 (1985)357} 
where the same 
result is advocated. We can use the result for $\ga^{(2)}$, Eq.~\Au{b}, 
to find $\beta^{(3)}_g$:
\eqn\russab{\eqalign{\lllf\beta^{(3)}_g =& 
4g^7 Q C(G)^2 -4g^5 C(G) r^{-1} \lf \Tr\left[ \ga^{(1)}C (R)\right] 
\cr&-2g^3 r^{-1} \llf \Tr\left[ \ga^{(2)}C (R)\right].\cr}}
Later we will use our result for $\ga^{(3)}$ to deduce $\beta^{(4)}_g$.

It follows from Eqs.~\russab, \Ed, \Ee{}\ and \Au{b} that 
in the $P={1\over3}Q$ case we have
\eqn\russc{\beta_g^{(3)} = {4\over {9}}{{g^7 Q^3}\over{(\lf )^3}}.} 
Our hypothesis is that a scheme
exists in which our result for $\ga^{(3)}$ 
satisfies Eq.~\Ena, and so Eq.~\russc\ then implies
\eqn\russd{\ga^{(3)} = {4\over {27}}{{g^6 Q^3}\over{(\lf )^3}}.}
We should emphasise that $\beta_g^{(3)}$  is sensitive to  coupling 
constant redefinitions of the form $\de g \sim O(g^5), (g^3 Y^2)$ etc, 
but {\it not\/} to redefinitions $\de Y$ of the same order: such 
redefinitions of $Y$ we will employ later.

There is little doubt that we should use the MS  results at one
and two loops on the right-hand side of Eq.~\russab.
However, at higher order, it is not immediately clear in
which scheme Eq.~\russa\ will be valid. The obvious expectation would 
be that MS is  appropriate. On the other hand, it is 
claimed 
\ref\lpsk{C.~Lucchesi, O.~Piguet and K.~Sibold, \plb 201 (1988) 241\semi 
C. Lucchesi, hep-th/9510078\semi
A.V.~Ermushev, D.I.~Kazakov and O.V.~Tarasov, \npb 281 (1987) 72}\ that there
are schemes in which a one-loop finite theory is finite to all orders; 
clearly such a scheme would not correspond to MS or \msbar.  Since as we 
already mentioned, Eq.~\russa\ gives $\beta_g^{(n+1)} =0$ 
for a $n$-loop finite theory, it might
be natural to believe that it would be a scheme of this type in which 
Eq.~\russa\ holds.
From our point of view the best-case scenario would be that there was a scheme
in which Eq.~\Ena\ and Eq.~\russa\ both held to all orders. 
We could then solve for
both $\beta_g$ and $\ga$, obtaining
\eqn\russb{
 \ga^i{}_j = {1\over{3g}}\beta_g\delta^i{}_j={{g^2}\over{\lf}}
\left[{Q\over{3 + 2 Q g^2 (\lf)^{-1}}}\right]\de^i{}_j .}
We then see that such a scheme would also be one for which a one-loop finite
theory is all-orders finite. 

\newsec{Three-loop finiteness}

In this section we review Parkes's result for $\ga^{(3)}$ in  a general
one--loop finite theory, and  show how a coupling constant redefinition 
reveals a finite theory. Thus throughout this section we 
shall set $P = Q = 0$. The result for $\ga^{(3)}$ in this special case,  
we will call $\ga_P^{(3)}$, is given by:  
\footnote{\dag}{We use the usual particle 
physicist's definition of the gauge coupling $g$; 
to compare with Ref.~\parkes\ 
one must set $g = 1/\sqrt{2}$} 
\eqn\tlfa{\eqalign{\lllf\ga^{(3)}_P = 
& \kappa g^6 \left[ 12 C(R) C(G)^2 - 2 C(R)^2 C(G) -10 C(R)^3   
-4 C(R)\Delta (R)\right]\cr   &
+ \kappa g^4\left[ 4C(R) S_1 - C(G)S_1 + S_2 -5 S_3\right] 
+\kappa g^2Y^* S_1 Y + \kappa M/4\cr}} 

where 
\eqna\tlfb$$\eqalignno{
S_1 ^i{}_j &= Y^{imn}C(R)^p{}_m Y_{jpn}&\tlfb a\cr
(Y^*S_1 Y)^i{}_j &= Y^{imn}S_1{}^p{}_m Y_{jpn}&\tlfb b\cr
S_2 ^i{}_j &= Y^{imn}C(R)^p{}_m C(R)^q{}_n Y_{jpq}&\tlfb c\cr
S_3 ^i{}_j &= Y^{imn}(C(R)^2)^p{}_m Y_{jpn}&\tlfb d\cr
M^i{}_j &= Y^{ikl}Y_{kmn}Y_{lrs}Y^{pmr}Y^{qns}Y_{jpq}&\tlfb e\cr
\Delta (R) &= \sum_{\alpha} C(R_{\alpha})T(R_{\alpha})&\tlfb f\cr}
$$
and $\kappa = 6\zeta (3)$. In Eq.~\tlfb{f}\ the sum over $\alpha$ is a sum 
over irreducible  representations. Thus whereas $C(R)$ is a matrix,
$C(R_{\alpha})$ and $\Delta (R)$ are numbers. 

Now let us consider the effect on $\ga_P$ of 
coupling constant redefinitions. 
Under a coupling constant redefinition of the form 

\eqn\tlfba{
Y^{ijk} \to Y^{ijk} + \delta Y^{ijk} \quad\hbox{and}\quad 
g \to g + \delta g}

we have 

\eqn\tlfbb{
(\delta\ga_P)^i{}_j = -\half ( Y^{ikl} \delta Y_{jkl} + \delta Y^{ikl} Y_{jkl}
 -8 g C(R)^i{}_j \delta g )}

We can exploit the freedom to make such coupling redefinitions to 
change $\ga_P^{(3)}$ so that it vanishes in a one--loop finite theory, 
as follows. Let us choose 
\eqn\tlfc{\eqalign{
(16\pi^2)^3\delta Y^{ijk} &=   k_1 g^2S_1{}^{(i}{}_m Y^{jk)m}
+ g^4[k_2C(R)^{(i}{}_m C(R)^j{}_n Y^{k)mn}\cr 
&+ k_3 Y^{n(jk}C(R)^{i)}{}_m C(R)^m{}_n + k_4\Delta (R) Y^{ijk}\cr 
&+ k_5 C(G) C(R)^{(i}{}_m Y^{jk)m} + k_6 C(G)^2 Y^{ijk}]\cr 
&+ k_7 Y^{ilm}Y^{jpq}Y^{krs}Y_{lpr}Y_{mqs}\cr
}}
(Note that our notation for symmetrising over $(ijk)$ involves a 
sum over three terms only: see Eq.~\Ec.)
The result is 
\eqn\tlfd{\eqalign{ 
(16\pi^2)^3\delta\ga_P = & -2 k_1 g^2( Y^* S_1 Y +2 g^2 C(R)S_1 + PS_1) 
-k_2 g^4( S_2 + 2 C(R)S_1)\cr
&-2 k_3g^4 ( S_3 + 2 g^2 C(R)^3 + PC(R)^2)
  - k_4 g^4\Delta (R)\left[ 2P + 4 g^2 C(R)\right] \cr 
&- 2 k_5 g^4C(G) \left[ S_1 + 2 g^2 C(R)^2  + PC(R)\right] 
- k_6g^4 C(G)^2 \left[2P + 4g^2C(R)\right]\cr 
&-k_7 M,
\cr}}
where, for later reference, we have temporarily reinstated factors of $P$. 
Then setting $P=0$ and 
\eqn\tlfdd{\eqalign{ 
k_1 = \ha\kappa, \qquad k_2 = &\kappa, \qquad k_3 = -{5\over2}\kappa,
 \qquad k_4 = - \kappa, \cr
\qquad k_5 = -\ha\kappa, 
\qquad k_6 =&  3\kappa, \qquad k_7 = {1\over 4}\kappa\cr}} 
we obtain $\delta\ga_P = - \ga_P^{(3)}$. Thus in an arbitrary one--loop
finite  theory one can transform $\ga_P^{(3)}$ to zero. This simple
demonstration from  Parkes's result seems to have eluded some previous
authors
\footnote{\dag}{Including, it must be said, one of the 
present authors\ref\tj{D.R.T.~Jones \npb 277 (1986) 153}} 
who have verified that it can be done in specific cases. 
There exist arguments 
that this procedure may in fact be extended to all orders\lpsk;
a specific calculation 
is always comforting, however, and in fact in some of the 
arguments presented in Refs.~\lpsk\ 
it is not entirely clear whether there is any constraint necessary 
with regard to the number of fields 
vis-\`a-vis the number of independent couplings. At least at three loops, 
we now see that there is none. (It is important that by  
one--loop finiteness we mean $\ga^{(1)} = \beta_g^{(1)} = 0$; it is easy to 
construct theories such that $\beta_Y^{(1)} = \beta_g^{(1)} = 0$, which 
are not even two--loop finite, and cannot be rendered so by coupling constant 
redefinitions.)    

We should note that for the finite theory,
the effect on $\ga^{(3)}$ of a redefinition $\delta g=a g$ is the same as 
that of
the redefinition $\delta Y^{ijk}=-aY^{ijk}$ (on use of the condition $P=0$),
and hence we could replace the $k_4$ and $k_6$ terms 
in Eq.~\tlfc\ with  a transformation on $g$ of the 
form 
\eqn\tlfda{
\delta g =g^4\left[k_4\Delta (R)+
k_6 C(G)^2\right].}
This would, however, change $\beta_g^{(3)}$; in particular we would 
no longer have $\beta_g^{(3)} = 0$ in $N=2$ theories (except, of course, 
for one--loop finite ones). We will return 
to the special case of $N =2$ in section 6.

We have concentrated above on the effect of coupling constant redefinitions 
on $\ga_P$; what is the effect of the corresponding redefinition on 
the $\beta$--function $\beta_Y$? In general we have to leading order in 
$\de Y, \de g$ that 
\eqn\tlfe{ \delta\beta_Y = \left[\beta_Y. {{\pa}\over{\pa Y}} 
+ \beta_Y^* . {{\pa}\over{\pa Y^*}} 
  + \beta_g . {{\pa}\over{\pa g}}\right]\delta Y 
- \left[ \delta Y . {{\pa}\over{\pa Y}} 
+ \delta Y^* . {{\pa}\over{\pa Y^*}}
+ \delta g . {{\pa}\over{\pa g}}\right] \beta_Y}
with a similar formula for $\delta\beta_g$. 
It is easy to see from this result 
that in a one loop finite theory we have simply   
\eqn\tlff{ \delta\beta_Y = Y^{m(ij}\delta\ga_P{}^{k)}{}_m,} 
since at one loop we have $\beta_Y = \beta_g = 0$. 
Hence if  $\ga_P^{(3)} + \de\ga_P = 0$ then $\beta^{(3)}_Y + \de\beta_Y = 0$
likewise.  It should be clear, however, that if we do not have $P = Q =
0$, the situation  changes and that after a coupling constant
redefinition the new $\beta_Y$ and  the new $\ga$ are not necessarily
related by Eq.~\Ec.  We will return to this point in section 7. 

\newsec{${\bf\ga^{(3)}}$ for a general theory}

Here we present the result for $\ga^{(3)}$ for a general non--abelian
theory.  The calculation is a straightforward application of the
superfield  Feynman rules spelled out in  Ref.~\ref\grsb{M.~Grisaru,
W~Siegel and M.~Rocek, \npb 159 (1979) 429} and applied to 
complementary calculations in Refs.~\parkes, \grs\ and \cz. (See also 
Ref.~\west, where $\ga^{(2)}$ is
calculated  for a general $N=1$ theory). In the $N=4$ case the set of 
Feynman diagrams to be calculated are to be found in Fig.~5 and  Fig.~6
of Ref.~\grs; we have to add diagrams with one or more  one-loop
self-energy insertions, and also a set of non-planar  diagrams which
happen to have vanishing group theory factors in  $N=4$ (Parkes
calculated the latter graphs too).   

One way in which we differ somewhat from some early calculations is
that we have performed the calculation  in the Feynman gauge. In the
literature one finds statements to the effect  that it is more
convenient or even essential to restore the  radiatively corrected 
gauge boson propagator to Feynman gauge form by redefining the gauge 
parameter. If this is done then the seagull graphs of the type  shown in
Fig.~7(b) of Ref.~\grs\ are zero; for us, however these  are non-zero,
but the graphs of the type Fig.~6(c) of the same reference are zero
instead,  because of the transverse nature of the gauge boson self
energy. We have  in fact checked that both procedures lead to the  same
result, but it seems to us simpler to stick to the Feynman gauge.  We 
evaluate the Feynman integrals by setting the external momentum zero,
and  introducing masses as necessary to control infra--red divergences.
We then  perform subtractions at the level of the Feynman integrals.
(A similar procedure is described in Ref.~\gmz.)
This  procedure gives an unambiguous result. No explicit factors of $d$
arise  in the algebra; this is important since such $d$-dependence would
require  careful handling. (For a discussion, see for example
Ref.~\ref\aj{R.~Allen and  D.R.T.~Jones, \npb 303 (1988) 271}). As 
explained in more detail later, we did not use the special case 
of $N=2$ \sy\ except as a final check, but we did use some 
of the graph--by--graph results from Ref.~\grs\ for $N=4$ 
to avoid particularly tedious calculations.  

The result for $\ga^{(3)}$ in a general theory is:  
\eqn\aga{\eqalign{\lllf\ga^{(3)} &= \lllf\ga_P^{(3)}\cr 
&+ \kappa \bigl\{ g^2\left[C(R)S_4 -2S_5 -S_6 \right] - g^4
\left[PC(R)C(G) +5PC(R)^2\right]\cr& 
+4g^6QC(G)C(R) \bigr\}
 +2Y^*S_4 Y - \half S_7 - S_8 +g^2\left[ 4C(R)S_4  + 4S_5\right]\cr
& + 
g^4\left[8C(R)^2 P  -2Q C(R) P - 4QS_1 
- 10r^{-1}{\rm Tr}\left[PC(R)\right]C(R)\right]\cr 
&  +g^6\left[2Q^2C(R)-8C(R)^2 Q  + 10QC(R)C(G)\right]
 \cr}}
where
\eqna\agab$$\eqalignno{
S_4 ^i{}_j &= Y^{imn}P^p{}_m Y_{jpn}&\agab a\cr
S_5^i{}_j &= Y^{imn}C(R)^p{}_m P^q{}_p Y_{jnq}&\agab b\cr
S_6 ^i{}_j &= Y^{imn}C(R)^p{}_m P^q{}_n Y_{jpq}&\agab c\cr
S_7 ^i{}_j &= Y^{imn}P^p{}_m P^q{}_n Y_{jpq}&\agab d\cr
S_8 ^i{}_j &= Y^{imn}(P^2)^p{}_m Y_{jpn}&\agab e\cr
(Y^*S_4 Y)^i{}_j &= Y^{imn}S_4{}^p{}_m Y_{jpn}.&\agab f\cr}
$$
Eq.~\aga\ is our main result. 
It is easy to see that for $P = Q = 0$ it  reduces to Parkes's result, 
Eq.~\tlfa. 

We can use this result for $\ga^{(3)}$ to write down $\beta_g^{(4)}$:
\eqn\agac{\eqalign{\llllf\beta_g^{(4)} &= 
8g^9 Q C(G)^3 -8g^7 C(G)^2 r^{-1} \lf \Tr\left[ \ga^{(1)}C (R)\right] 
\cr&-4 C(G) g^5 r^{-1} \llf \Tr\left[ \ga^{(2)}C (R)\right]
-2g^3 r^{-1} \lllf\Tr\left[\ga^{(3)}C(R)\right].\cr}}
Naturally the precise result for $\beta_g^{(4)}$ depends on the scheme 
that we use to evaluate $\ga^{(3)}$; as we already indicated, it is 
probably appropriate to use MS, that is Eq.~\aga.  

Let us compare our result for $\ga^{(3)}$ with previous calculations.
The simplest possible case is the Wess-Zumino model, corresponding
to $g = 0$ and a superpotential $W = {1\over 6}\la\Phi^3$. There are
several results to choose from in the literature; the original
calculation \ref\abgr{L.F.~Abbott and M.T.~Grisaru, \npb 169 (1980) 415},
and two subsequent efforts
\ref\sen{A.~Sen and M.K.~Sundaresan,  \plb 101 (1981) 61}
\ref\avdeev{L.V.~Avdeev et al, 
\plb 117 (1982) 321}, both of which in fact proceeded to four loops.
The three calculations differ with regard to the coefficient of the 
$\zeta (3)$ term at three loops. 

It is easy to show that for the Wess--Zumino model our result is
\eqn\agaca{
\ga = \ha \left({\la\over{4\pi}}\right)^2
-\ha\left({\la\over{4\pi}}\right)^4
+ \left( {5\over 8} +{3\over 2}\zeta (3)\right)
\left( {{\la}\over{4\pi}}\right)^6 + \cdots}
and that this agrees with Ref.~\avdeev.

Another interesting check on our result is afforded by
the case of $N =2$ \sy. We discuss this in the next section and then 
consider the effect of coupling constant redefinitions on Eq.~\aga, 
with emphasis on theories that satisfy $P = {1\over 3}Q$.

\newsec{The ${\bf N = 2}$ case}

In $N = 1$ language, an $N =2$ theory is defined by the superpotential 
\eqn\agb{
W = \sqrt{2}g \eta^a \chi^i S^a{}^j{}_i \xi_j
}
where $\eta, \chi$ and $\xi$ transform according to the adjoint, $S^*$
and $S$  representations respectively. The set of chiral superfields 
$\chi, \xi$  is called a hypermultiplet. (In the abelian case
$\eta$ is neutral and   $\chi, \xi$ are two sets of fields in opposite
charge pairs. Abelian $N=2$ without  hypermultiplets is a free field
theory; but with them we have non-trivial interactions). 
$N=2$ theories have one loop divergences only\ref\hsw{
P.S.~Howe, K.S.~Stelle and P.~West, \plb 124 (1983) 55\semi
P.S.~Howe, K.S.~Stelle and P.K.~Townsend, \npb  236 (1984) 125}; 
using MS or \msbar\ 
we may therefore expect that the anomalous  dimension of both 
the $\eta$ and the hypermultiplet should vanish beyond one loop.
Parkes, in fact, used this result to reduce calculational labour; 
we have preferred to use it as a check. 

At the one--loop level we have 
\eqn\agc{\eqalign{Q &= 2 \left[T(S) - C(G)\right]\cr
P_{\eta} &= Q\de^a{}_b\cr
P_{\chi} &= P_{\xi} = 0.\cr}}
Thus except for the case $P = Q = 0$, $N=2$ theories 
cannot satisfy $P = {1\over 3}Q$.

In Table 1 we give expressions for group theory factors defined in 
Eqs.~\tlfb{}\ and \agab{}, when specialised to $N=2$; using these results
we may readily demonstrate that 
Eq.~\aga\ vanishes identically for both $\eta$  and the 
hypermultiplet. We note also in passing that the exact $\beta$ function result
in Eq.~\russa\ is consistent with neither $\beta_g$ nor $\ga$ receiving 
corrections beyond one loop in the $N=2$ case. 
\vfill\eject
$$\vbox{\offinterlineskip
\def\vr{\vrule height 11pt depth 5pt}
\def\vrq{\vr\quad}
\settabs
\+
\vrq Group theory factor  \quad & \vrq \qquad 
Contribution to $\eta$ \quad\qquad\  
& \vrq \quad Contribution to $\xi, \chi$ \quad\quad\ &
\vr\cr\hrule
\+
\vrq Group theory factor  \quad & \vrq \quad 
Contribution to $\eta$ \quad\quad  
& \vrq \quad Contribution to $\xi, \chi$ \quad\quad\ &
\vr\cr\hrule
\+
\vrq $C(R)S_1$  \quad & \vrq  \quad $4C(G)\Delta (S)$
& \vrq $2C(S)^2\left[C(S)+C(G)\right]$\quad&
\vr\cr\hrule
\+
\vrq $C(G)S_1$  \quad & \vrq  \quad $4C(G)\Delta (S)$
& \vrq $2C(S)C(G)\left[C(S)+C(G)\right]$\quad&
\vr\cr\hrule
\+  
\vrq $QS_1$  \quad & \vrq  \quad $4Q\Delta (S)$
& \vrq $2QC(S)\left[C(S)+C(G)\right]$\quad&
\vr\cr\hrule  
\+ 
\vrq $Y^*S_1Y$  \quad & \vrq   $8C(G)\Delta (S)$
& \vrq $4\left[C(S)+C(G)\right]C(S)^2$\quad&
\vr\cr 
\+
\vrq  \quad & \vrq  \quad $+8\sum \left[C(S_{\alpha})\right]^2T(S_{\alpha})$
& \vrq $+8C(S)\Delta (S)$\quad&
\vr\cr\hrule
\+
\vrq $S_2$  \quad & \vrq  \quad 
$4\sum \left[C(S_{\alpha})\right]^2T(S_{\alpha})$
& \vrq $4C(G)C(S)^2$\quad&   
\vr\cr\hrule   
\+
\vrq $S_3$  \quad & 
\vrq  \quad $4\sum \left[C(S_{\alpha})\right]^2T(S_{\alpha})$
& \vrq $2\left[C(S)^2+C(G)^2\right]C(S)$\quad&
\vr\cr\hrule
\+
\vrq $C(R)S_4$  \quad & \vrq  \quad 0
& \vrq $2QC(S)^2$\quad&
\vr\cr\hrule
\+ 
\vrq $Y^*S_4Y$  \quad & \vrq  \quad $8Q\Delta (S)$
& \vrq $4QC(S)^2$\quad&
\vr\cr\hrule  
\+
\vrq $S_5$  \quad & \vrq  \quad 0
& \vrq $2QC(G)C(S)$\quad&             
\vr\cr\hrule
\+
\vrq $S_6$  \quad & \vrq  \quad 0    
& \vrq $2QC(S)^2$\quad&         
\vr\cr\hrule
\+
\vrq $S_7$  \quad & \vrq  \quad 0  
& \vrq 0\quad&          
\vr\cr\hrule
\+
\vrq $S_8$  \quad & \vrq  \quad 0  
& \vrq $2Q^2C(S)$\quad&         
\vr\cr\hrule
\+
\vrq $C(R)\Delta (R)$  \quad & 
\vrq  \quad $C(G)\left[ C(G)^2 + 2\Delta (S)\right]$
& \vrq $C(S)\left[C(G)^2 + 2C(S)\Delta (S)\right]$\quad&
\vr\cr
\hrule
\+
\vrq $M$  \quad & \vrq   $16C(G)\left[C(G)T(S)-3\Delta(S)\right]$
& \vrq $16[C(G)^2-3C(S)C(G)$\quad&
\vr\cr
\+
\vrq   \quad & \vrq  \quad $+32\sum C(S_{\alpha})^2T(S_{\alpha})$       
& \vrq $+2C(S)^2]C(S)$\quad&
\vr\cr\hrule
\+
\vrq $C(R)^2P$  \quad & \vrq  \quad $C(G)^2Q$
& \vrq 0\quad&
\vr\cr\hrule
\+  
\vrq $C(R)^2Q$  \quad & \vrq  \quad $C(G)^2Q$
& \vrq $C(S)^2Q$\quad&         
\vr\cr\hrule
\+
\vrq $QC(R)P$  \quad & \vrq  \quad $Q^2C(G)$
& \vrq 0 \quad&
\vr\cr\hrule
\+
\vrq $Q^2C(R)$  \quad & \vrq  \quad $Q^2C(G)$
& \vrq $Q^2C(S)$\quad&
\vr\cr\hrule
\+
\vrq $\Tr[PC(R)]C(R)$  \quad & \vrq  \quad $QC(G)^2$
& \vrq $QC(G)C(S)$\quad&
\vr\cr\hrule
}$$

\in
{\it \noindent 
Table 1: Group theory factors for the special case of $N=2$ supersymmetry}
\out

\newsec{Coupling  constant redefinitions and ${\bf P= {1\over 3}Q}$}

We now wish to investigate whether our three-loop result for $\ga^{(3)}$
can be 
transformed by coupling constant redefinition into the conjectured formula 
Eq.~\russd\ in the case where we impose Eq.~\El,
in accord with our hypothesis. Clearly the redefinition must at
least include the redefinition of Eq.~\tlfd\ in order to remove the terms in
$\ga_P^{(3)}$. If we make this redefinition, then we obtain  
\eqn\ccra{\eqalign{\lllf(\ga^{(3)} +\delta\ga_P)&=  
\kappa \bigl\{ g^2\left[C(R)S_4 -2S_5 -S_6 -PS_1\right] 
+2g^4 P \left[ \Delta (R) -3 C(G)^2\right]\cr&
+4g^6QC(G)C(R) \bigr\}
 +2Y^*S_4 Y - \half S_7 - S_8 \cr
&+g^2\left[ 4C(R)S_4  + 4S_5\right]\cr
& +
g^4\left[8C(R)^2 P  -2Q C(R) P - 4QS_1
- 10r^{-1}{\rm Tr}\left[PC(R)\right]C(R)\right]\cr
&  +g^6\left[2Q^2C(R)-8C(R)^2 Q  + 10QC(R)C(G)\right],
 \cr}}
where we have not yet imposed Eq.~\El. 
Note that we have now included the terms 
involving $P$ in Eq.~\tlfd.
If we now impose $P={1\over3}Q$, 
then the result simplifies dramatically:
\eqn\ccrb{\eqalign{\lllf(\ga^{(3)}+\delta\ga_P)
&=\kappa\Bigl\{-{4\over3}g^4 QS_1
+g^6\Bigl[{4\over3}QC(R)^2+4QC(R)C(G)\cr
&+{2\over 3}Q\Delta (R) - 2QC(G)^2
+{2\over9}Q^2C(R)\Bigr]\Bigr\}\cr
&  +{5\over 27}g^6Q^3.\cr}}
 We can now try to redefine $\ga^{(3)}$ still further in order to remove the
terms proportional to $\kappa$. Let us focus our attention on the terms 
involving $QS_1$ and $QC(R)^2$. There is one coupling constant 
redefinition which changes the coefficients of both these terms
simultaneously, namely
\eqn\ccrc{
\lllf\delta Y^{ijk}=k_8g^4QC(R)^{(i}{}_lY^{jk)l}.}
The corresponding change in $\ga^{(3)}$ is given according to Eq.~\tlfbb\ by
\eqn\ccrd{
\lllf\delta\ga_Q= -2k_8Qg^4\left[S_1+2g^2C(R)^2+{1\over3}QC(R)\right].}
There is no other redefinition which affects the coefficients of either 
$QS_1$ or $QC(R)^2$. 
Since the coefficients of $QS_1$ and $QC(R)^2$ are in a different
ratio in Eq.~\ccrb\ compared with Eq.~\ccrd, 
we cannot simultaneously redefine 
both of them to zero. Hence we cannot remove all the terms proportional to
$\kappa$ in $\ga^{(3)}$, and so our conjectured result Eq.~\russd\ cannot be 
true in any renormalisation scheme. 
Moreover, the fact that neither $QS_1$ nor
$QC(R)^2$ is in general proportional to the unit matrix means that Eq.~\Ena\
cannot be true at three loops whatever the value of $\beta_g^{(3)}$.
 However, in principle, we could remove all the 
remaining terms proportional to $\kappa$ by a redefinition of the form
\eqn\ccrda{\eqalign{\lllf\delta Y^{ijk}&=ag^4Y^{ijk},\cr
                    \lllf\delta g&=bg^5,\cr}}
which produces a change in $\ga$ of the form
\eqn\ccrdb{\lllf\delta \ga=4(b-a)g^6C(R)-{2\over3}g^6aQ.} 
We should note, though,  that the redefinition of $g$ would in principle be 
fixed by  
a three-loop calculation of $\beta_g$. A redefinition $\delta g=bg^5$  
produces a change
$\delta\beta_g=2bQg^7$ in $\beta_g$. After computing $\beta_g$ at three 
loops in MS, $\delta g$ would be fixed by requiring that it 
transform the result to the form corresponding to Eq.~\russc. 
(Since $\beta_g^{(1)}$ contains
no $Y$-dependence, redefinitions of $Y$ of the form we are considering
in Eqs.~\tlfc\ and  \ccrc\ do not change $\beta_g$ at three loops.)
Of course it might be the case that Eq.~\russa, and hence Eq.~\russc, is true 
already in minimal 
subtraction, in which case no redefinition of $g$ would be required.

We now return to the ideas of coupling constant reduction, which we briefly
described earlier. Eqs.~\Ema\ and \Emc\ may be rewritten as
\eqn\altb{Y^{\prime ijk}\equiv\lambda^{\prime ijk}(g)=ga^{ijk},}
where
\eqn\altc{\lambda^{\prime ijk}=\lambda^{ijk}-g^3 b^{ijk} -g^5 c^{ijk}
-\ldots.}
Eq. ~\altb\ implies
\eqn\altd{\beta_{Y'}^{ijk}= g^{-1}Y^{\prime ijk}\beta_g.}
In other words, if the CCR program 
can be carried out, there is a renormalisation scheme in which the
redefined coupling $Y'\over g$ is at a fixed point.
In our case, we could imagine implementing the CCR 
program by starting with a solution $Y^{ijk}=ga_1^{ijk}$ of Eq.~\El\ 
as in Eq.~\Elbb\ and adding corrections to
obtain an RG invariant relation as in Eq.~\Emc.
After redefining $\lambda^{ijk}$ (and hence $Y^{ijk}$) 
as in Eq.~\altc, the new coupling would
solve $P={1\over3}Q$ and would satisfy Eq.~\Elb. 
Hence we would want to find a scheme 
in which Eq.~\Elb\ is true upon imposing $P={1\over3}Q$.
Our failure 
to achieve Eq.~\Ena\ for a general theory with arbitrary coupling constant 
redefinitions might already make us suspect that 
we will also be unable to obtain Eq.~\Elb\ in general. Let us check this in 
detail. 
We readily find from Eqs.~\Ec\ and \tlfe\ that the effect of changes
$\delta Y^{ijk}$ and $\delta g$ of the general form Eqs.~\tlfc\ or \ccrc\
upon the
$\beta$-function for $Y^{ijk}$ is given in the $P={1\over3}Q$ case by
\eqn\ccre{\delta\beta_Y^{ijk}=4Qg^2\delta Y^{ijk}+Y^{m(ij}\delta
\ga^{k)}{}_m,}
where $\delta\ga$ is given by Eq.~\tlfbb. Clearly the redefinition we make
must at least contain $\delta\ga_P$
corresponding to substituting Eq.~\tlfdd\ in Eq.~\tlfd. If we make this 
redefinition alone,
we find that $\beta_Y$ is transformed to
\eqn\ccrf{\beta_Y^{\prime ijk}=\beta_Y^{ijk}+\delta\beta_Y^{ijk}=4Qg^2
\delta Y^{ijk}+Y^{m(ij}\ga^{\prime k)}{}_m,}
where
\eqn\ccrg{\eqalign{
(16\pi^2)^3\delta Y^{ijk} &=   
\kappa\bigl\{ {1\over2}g^2S_1{}^{(i}{}_m Y^{jk)m}
+ g^4\bigl[C(R)^{(i}{}_m C(R)^j{}_n Y^{k)mn}\cr
&-{5\over2}Y^{n(jk}C(R)^{i)}{}_m C(R)^m{}_n -\Delta (R) Y^{ijk}\cr   
&-{1\over2}C(G) C(R)^{(i}{}_m Y^{jk)m} + 3C(G)^2 Y^{ijk}\bigr]\cr
& + {1\over4}Y^{ilm}Y^{jpq}Y^{krs}Y_{lpr}Y_{mqs}\bigr\}\cr}}
and
\eqn\ccrh{
\ga^{\prime i}{}_j=\ga^i{}_j+\delta\ga_P^i{}_j={1\over{3g}}(\beta_g^{(1)}
+\beta_g^{(2)}+\beta_g^{(3)})\delta^i{}_j+Qg^2X^i{}_j}
with
\eqn\ccrha{\eqalign{\lllf X&=\kappa \Bigl\{-{4\over3}g^2S_1
\cr&+g^4\bigl[
4C(R)C(G)+{4\over3}C(R)^2\cr
&+{2\over9}QC(R)+{2\over3}\Delta(R)-2C(G)^2
\bigr]\Bigr\}+{1\over{27}}Q^2.\cr}}
Clearly we can write Eq.~\ccrf\ in the form, correct to three loops,
\eqn\ccri{\beta_Y^{\prime ijk}=g^{-1}\beta_gZ^{ijk},}
where
\eqn\ccrj{Z^{ijk}=Y^{ijk}+4\delta Y^{ijk}+Y^{m(ij}X^{k)}{}_m.}
The fact that $Z^{ijk}\ne Y^{ijk}$ means that we have failed to provide 
an explicit construction of the 
CCR program in this context. It is not difficult to 
see that further redefinitions will not save the day. 
In particular, there is no
way to cancel the $Y^5$ term in $\delta Y$, which was required to remove the 
$M$ term in $\ga^{(3)}$. 

Now  according to Ref.~\zimm, the existence of a  one--loop CCR
construction is sufficent to establish it to all orders.  
If we accept this, then clearly the existence of $a_1^{ijk}$ 
satisfying  Eq.~\Elbb\ assures us that there exists a
fixed point of $Y^{ijk}/g$ for a suitably redefined $Y$.   It is
disappointing that this result does not emerge naturally  in our
formalism (in contrast to the way we found the explicit  redefinition
that renders a one--loop  finite theory three loops finite). At three 
loops we  cannot find a general expression for the required coupling
constant redefinition in terms of $Y^{ijk}$ and $C(R)^i{}_j$ (or
equivalently we cannot  construct the expansion Eq.~\Emc\ for a general
theory). This does not mean that the redefinition does not exist in 
special cases; but we must rely on Ref.~\zimm\ in order to assert 
that it is a general result.

\newsec{Conclusions}

Our main new result here is the MS expression for the three--loop  chiral
superfield  anomalous dimension in a general $N = 1$ theory, Eq.~\aga.
Since  in  Eq.~\russab\ we also gave the   three--loop gauge 
$\beta$--function, we  have the complete set of 
$\beta$--functions for dimensionless couplings in an arbitrary 
$N=1$ theory. It will be interesting to examine the effect 
of the consequent corrections to the standard running coupling 
analysis. 

Another  motivation  for the calculation was our previous observation
that a certain simple relation  between the Yukawa and gauge couplings
is RG invariant through two loops. We have found that even with
arbitrary  coupling constant redefinitions this property does not extend
to three  loops, in general, at least. That is to say, we are unable to
achieve   $\ga^{(3)}\sim \de^i{}_j$,  so that Eq.~\Ena\ cannot be true,
in general, irrespective of the  value of $\beta_g^{(3)}$. We have also
shown, however, that given a  theory satisfying the (one-loop) $P =
{1\over 3}Q$ condition, the CCR  paradigm assures us of the existence of
a fixed point to all orders.  This suffices to make such theories
phenomonologically interesting.

\bigskip\centerline{{\bf Acknowledgements}}\nobreak

IJ and CGN were supported by PPARC via an Advanced Fellowship and a 
Graduate Studentship respectively. 

\listrefs 

\bye